\documentclass[a4paper,twocolumn,english,aps,prl,superscriptaddress]{revtex4-1}
\usepackage[T1]{fontenc}
\usepackage[latin9]{inputenc}
\usepackage{color}
\usepackage{babel}
\usepackage{amsmath}
\usepackage{amssymb}
\usepackage{graphicx}
\usepackage[unicode=true,pdfusetitle,
 bookmarks=true,bookmarksnumbered=false,bookmarksopen=false,
 breaklinks=false,pdfborder={0 0 1},backref=false,colorlinks=false]
 {hyperref}

\makeatletter

\usepackage{babel}
\usepackage{babel}

\makeatother

\begin{document}
\renewcommand{\figurename}{FIG.} 
\title{Anomalous Andreev spectrum and transport in non-Hermitian Josephson
junctions}
\author{Chang-An Li}
\email{changan.li@uni-wuerzburg.de}

\affiliation{Institute for Theoretical Physics and Astrophysics, University of
Würzburg, 97074 Würzburg, Germany}
\author{Hai-Peng Sun}
\affiliation{Institute for Theoretical Physics and Astrophysics, University of
Würzburg, 97074 Würzburg, Germany}
\author{Björn Trauzettel}
\affiliation{Institute for Theoretical Physics and Astrophysics, University of
Würzburg, 97074 Würzburg, Germany}
\affiliation{Würzburg-Dresden Cluster of Excellence ct.qmat, Germany}
\date{\today}
\begin{abstract}
We propose a phase-biased non-Hermitian Josephson junction (NHJJ)
composed of two superconductors mediated by a short non-Hermitian
link. Such a NHJJ is described by an effective non-Hermitian Hamiltonian
derived based on the Lindblad formalism in the weak coupling regime.
By solving the Bogoliubov-de Gennes equation, we find that its Andreev
spectrum as a function of phase difference exhibits Josephson gaps,
i.e. finite phase windows with no Andreev (quasi-)bound states. The
complex Andreev spectrum and the presence of Josephson gaps constitute
particular spectral features of the NHJJ. Moreover, we propose complex
supercurrents arising from inelastic Cooper pair tunneling to characterize
the anomalous transport in the NHJJ. Additional numerical simulations
complement our analytical predictions. We demonstrate that the Josephson
effect is strongly affected by non-Hermitian physics. 
\end{abstract}
\maketitle

\section{Introduction}

Recently, there has been growing interest in non-Hermitian
systems\ \cite{Bender07RPP,El-Ganainy18nphys,Ashida20AP,Bergholtz21rmp,Okuma23arcmp}.
Non-Hermitian physics emerges in a variety of fields such as disordered
systems\ \cite{Zyuzin18prb,Papaj19prb}, correlated solids\ \cite{Yoshida19prb,Nagai20prl},
and open or driven systems\ \cite{Makris08prl,GuoA09prl,Rotter09JA,Nakagawa18prl,Xiaol19prl,Lij19nc,Zhangx20prb}.
Different from the Hermitian case, the eigenvalues of non-Hermitian
systems are complex in general. The complex-valued nature of the spectrum
gives rise to unique features such as exceptional points\ \cite{Dembowski01prl,Berry04CJP,Kawabata19prl},
point gaps\ \cite{Kawabata19prx,Zhou19prb}, and enriched topological
phases\ \cite{Lee16prl,Yao18prl,Kunst18prl,GonZP18prx}. Many of
these particular phenomena have been theoretically studied and experimentally
observed\ \cite{Lee16prl,Yao18prl,Kunst18prl,GonZP18prx,Zhou19prb,Kawabata19prx,Kawabata19prl,ZhangK20prl,Okuma20prl,Yao18prl2,ZengQB16pra,San-Jose16SR,Leykam17prl,ShenH18prl,Yin18prb,Yokomizo19prl,LeeCH19prb,Longhi19prl,LeeJY19prl,Yamamoto19prl,Avila19CP,Borgnia20prl,Budich20prl,Yamamoto21prl,Franca22prl,Cayao22prb,Jezequel23prl,QinF23pra,GuoC23prl,Cayao23prb,Sun23prb,LiCA23prl,Kornich23prl,Zeuner15prl,DingK16prx,Xiao20np,Fahri21science,LiangQ22prl,ZhangQ23prl,LiangC23prl,XuG23prl}.
Whereas the research on characterizing novel non-Hermitian systems
has achieved impressive progress, the impact of non-Hermiticity on
quantum transport is largely unexplored despite some related efforts\ \cite{ZhuB16pra,Longhiprb17,ChenY18prb,Bergholtz19prr,Shobe21prr,Michen22prr,Viktoriia22prr,Sticlet22prl,GengH23prb,Isobe23prb},
especially for the cases involving superconductivity. For instance,
an intriguing open question is how non-Hermiticity affects low-energy
spectral features of superconducting systems and results in particular
transport signatures. 

The best settings to study superconducting transport are Josephson
junctions (JJs). Generally, JJs consist of two superconductors separated
by a (weak) link\ \cite{Likharev79rmp,Beenakker92proceed,Golubov04rmp,Tinkham}.
Hermitian JJs have been extensively studied for decades. Phase-coherent
quantum transport in Hermitian JJs gives rise to the dc Josephson
effect: a phase-dependent supercurrent without voltage bias. In the
short junction regime, the Josephson effect is entirely determined
by the low-energy Andreev spectrum\ \cite{Furusaki99sm,Beenakker91prl2,Kwon04epj,FuL09prb,Dolcini15prb,Beenakker13prl}.
When non-Hermiticity is introduced, we expect that the impact of non-Hermiticity
on fundamental properties of JJs, such as the low-energy Andreev spectrum
and the phase-biased supercurrent, is substantial. It is also of experimental
relevance because JJs are always coupled to the environment and non-Hermitian
parts of a Hamiltonian mimic certain aspects of this coupling. 

In this paper, we investigate the low-energy spectrum and transport
properties of a short non-Hermitian Josephson junction (NHJJ). The
proposed NHJJ is built up of two superconductors connected by a non-Hermitian
barrier {[}Fig. \ref{fig1:setup}(a){]}. We discover that the Andreev
spectrum becomes complex-valued and exhibits Josephson gaps {[}purple
regions in Fig. \ref{fig1:setup}(b){]}. Due to the complex-valued
nature of the Andreev spectrum, the supercurrent becomes complex.
Its imaginary part signifies the loss of quasiparticles due to inelastic
Cooper pair tunneling. We also perform detailed tight-binding calculations
to simulate experimental implementations of our proposal. 

The paper is organized as follows. In Sec. II, we present the construction
of the NHJJ and its effective non-Hermitian Hamiltonian from the Lindblad
formalism in a weak coupling regime. In Sec. III, we derive the Andreev
quasi-bound states and develop the concept of Josephson gaps. In Sec.
IV, we present the inelastic Cooper pair tunneling and its consequent
complex supercurrents. In Sec. V, we consider the experimental feasibility
and numerical simulations of the NHJJ. Finally, we conclude in Sec.VI.

\section{Effective non-Hermitian Hamiltonian of the non-Hermitian Josephson
junctions}

\begin{figure}
\includegraphics[width=1\linewidth]{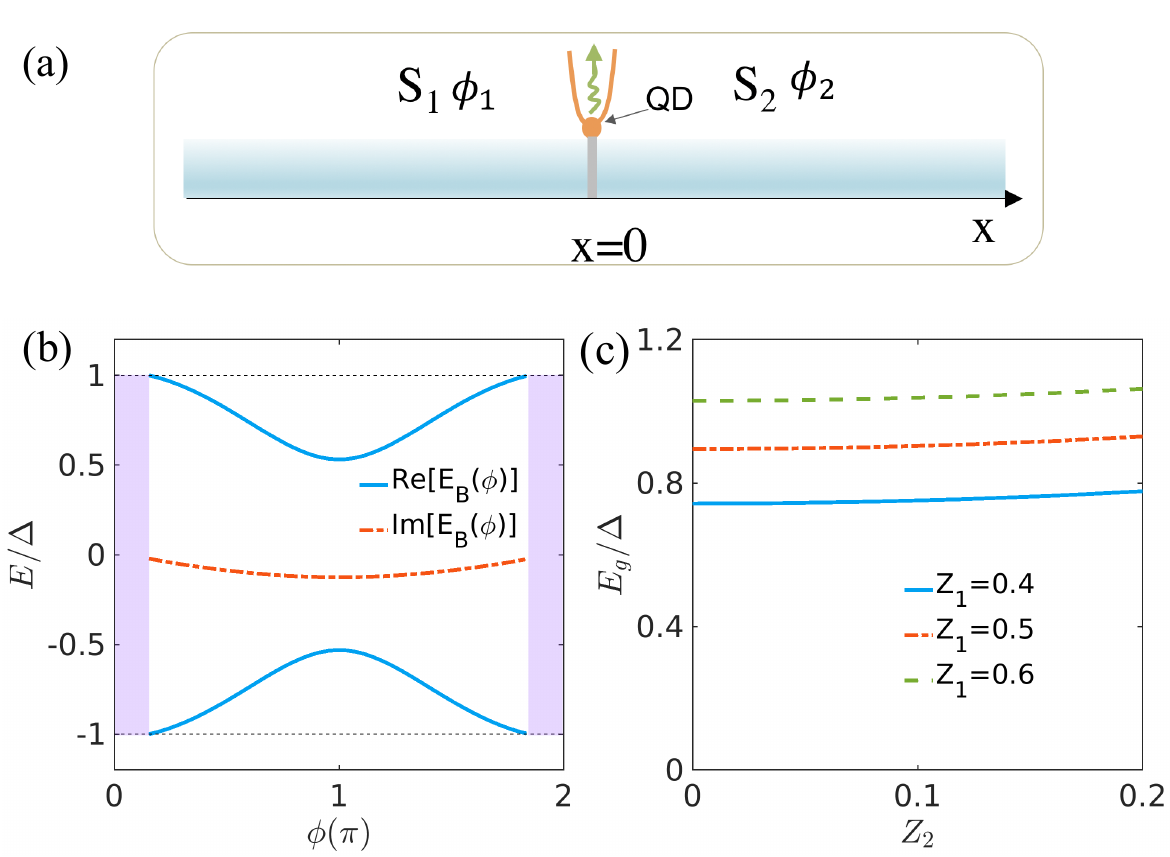}

\caption{(a) Sketch of a 1D NHJJ: two superconductors are connected by a short
junction, which is coupled to the environment (a reservoir and a quantum
dot (QD)). The QD is used to monitor the tunneling of quasiparticles
between system and reservoir. The two superconductors possess a phase
difference $\phi=\phi_{1}-\phi_{2}$. The wavy line indicates loss
due to the coupling to the environment. (b) Complex Andreev spectrum
as a function of $\phi$ within the superconducting gap $\Delta$
in the $s$-wave NHJJ. The purple regions indicate Josephson gaps.
We choose $Z_{1}=0.6$ and $Z_{2}=0.2$ here. (c) The energy gap $E_{g}$
of Andreev spectrum as a function of the parameter $Z_{2}$ for different
choices of $Z_{1}=0.4,0.5$ and $0.6$. \label{fig1:setup}}
\end{figure}

We consider a minimal model of a one-dimensional (1D) NHJJ as sketched
in Fig. \ref{fig1:setup}(a). The junction is assumed to be coupled
to a fermionic environment, constituting an open quantum system. We
derive an effective NH Hamiltonian description for this system from
the Lindblad formalism under certain approximations\ \cite{Breuer02book,Daley14AIP,Minganti19pras,Manzano20AIP,Roccati22OS}.
The Hamiltonian for the total quantum system is composed of three
parts 
\begin{equation}
H=H_{S}+H_{E}+H_{I},
\end{equation}
where $H_{S}$ is the system Hamiltonian, describing the JJ without
coupling to the environment. The degrees of freedom in the environment
are described by the Hamiltonian $H_{E}$. The interaction between
system and environment is captured by $H_{I}$. The Hamiltonian of
the system can be written in its eigenbasis as $H_{S}=\sum_{n_{s}}(\varepsilon_{n_{s}}-\mu_{s})C_{n_{s}}^{\dagger}C_{n_{s}}$
with eigenenergy $\varepsilon_{n_{s}}$ and eigenstate $|n_{s}\rangle$,
where $C_{n_{s}}^{\dagger}$ ($C_{n_{s}}$) denotes creation (annihilation)
operators of the system. They refer to the occupation of Andreev quasi-bound
states. The fermionic environment is described by the Hamiltonian
$H_{E}=\sum_{n_{e}}(\varepsilon_{n_{e}}-\mu_{e})C_{n_{e}}^{\dagger}C_{n_{e}}$.
We assume that $\mu_{e}<\mu_{s}$, i.e. the chemical potential of
the environment is smaller than that of the system. We envision that
an external reservoir is coupled to the JJ via a quantum dot, cf.
Fig. \ref{fig1:setup}(a). The quantum dot works as a monitor of the
tunneling particles between system and reservoir, as we elaborate
on below. Note that the Hamiltonian $H_{E}$ describes the degrees
of freedom of the external reservoir and the quantum dot. The coupling
Hamiltonian is given by $H_{I}=\sum_{n_{e},n_{s}}V_{es}C_{n_{e}}^{\dagger}C_{n_{s}}+h.c.$.
In general, the interaction term can be decomposed as $H_{I}=\sum_{i}S_{i}\otimes E_{i}$,
where $S_{i}$ ($E_{i}$) is an operator acting on the system (the
environment). We make a weak coupling assumption, in the sense, that
the coupling between system and environment is weaker than the energy
gap of Andreev levels. 

The Lindblad master equation describes the dynamics of the system
in the presence of dissipation due to the environment. The Lindblad master
equation takes the form\ \cite{Breuer02book,Daley14AIP,Minganti19pras,Manzano20AIP,Roccati22OS}

\begin{alignat}{1}
\frac{d\rho_{S}}{dt} & =-i[H_{S},\rho_{S}]+\sum_{m}\gamma(\hat{L}_{m}\rho_{S}\hat{L}_{m}^{\dagger}-\frac{1}{2}\{\hat{L}_{m}^{\dagger}\hat{L}_{m},\rho_{S}\}),\label{eq:LindbladEq}
\end{alignat}
where $\rho_{S}$ is the reduced density matrix of the system and
$\hat{L}_{m}$ the Lindblad operator describing the dissipative dynamics
(including decoherence and loss processes) occurring at a positive
rate $\gamma$\ \cite{Daley14AIP}. We can rewrite Eq.~\eqref{eq:LindbladEq}
as 

\begin{alignat}{1}
\frac{d\rho_{S}}{dt} & =-i(H_{\mathrm{eff}}\rho_{S}-\rho_{S}H_{\mathrm{eff}}^{\dagger})+\sum_{m}\gamma\hat{L}_{m}\rho_{S}\hat{L}_{m}^{\dagger},
\end{alignat}
where $H_{\mathrm{eff}}=H_{S}-\frac{i}{2}\sum_{m}\gamma\hat{L}_{m}^{\dagger}\hat{L}_{m}.$
Considering $\mu_{e}<\mu_{s}$, we assume that the energy modes of
the environment coupled to the system are unoccupied\ \cite{Daley14AIP}.
In this case, we have $\hat{L}_{m}=C_{m_{s}}$. 

The dissipation term $-\frac{i}{2}\sum_{m}\gamma\hat{L}_{m}^{\dagger}\hat{L}_{m}$
describes the continuous loss of coherence and energy of the system to
the environment. Without the quantum-jump term\ \cite{Yamamoto19prl,Minganti19pras,Roccati22OS},
we obtain the equation of motion $\frac{d\text{\ensuremath{\rho_{S}}}}{dt}=-i(H_{\mathrm{eff}}\rho_{S}-\rho_{S}H_{\mathrm{eff}}^{\dagger})$,
which is equivalent to work with an effective non-Hermitian Hamiltonian
\begin{equation}
H_{\mathrm{eff}}=H_{S}-\frac{i}{2}\sum_{m}\gamma\hat{L}_{m}^{\dagger}\hat{L}_{m}.
\end{equation}

The quantum-jump term $\sum_{m}\gamma\hat{L}_{m}\rho_{S}\hat{L}_{m}^{\dagger}$
is related to the abrupt change of quantum trajectories \cite{Minganti19pras,Daley14AIP}.
The Lindblad operator takes $\hat{L}_{m}=C_{m_{s}}$ in our case.
Consider a state $|m_{s}\rangle=|0,\cdots,n_{m},\cdots\rangle$ in
the system $H_{s}$. Then the quantum-jump term $\gamma C_{m_{s}}|m_{s}\rangle\langle m_{s}|C_{m_{s}}^{\dagger}$
indicates the annihilation of an excitation in the state $|m_{s}\rangle$.
This excitation escapes into the environment. To monitor the quantum-jump
process, we assume an auxiliary quantum dot between the junction and
the external reservoir, as shown in Fig. \ref{fig1:setup}(a). Note
that this setup can be experimentally realized \cite{Deng16Sciences}.
The quantum dot has a few energy levels. Hence, it is feasible to
monitor its occupation e.g. by a nearby quantum point contact. By
performing weak measurements and postselection techniques on the quantum
dot\ \cite{Naghiloo19NP}, it is possible to only consider those
trajectories where no quantum jumps happen during finite time windows.
The average over many such quantum trajectories describes the loss
physics due to the dissipation. 

To capture this loss physics of the system in a minimal way, we employ
a non-Hermitian potential at the junction as $-iV_{2}\delta(x)$ with
$V_{2}>0$. This non-Hermitian potential mimics the loss of coherence
and particles effectively. Furthermore, we assume that the junction
between two superconductors also exhibits a real barrier potential
described by $V_{1}\delta(x)$ with $V_{1}>0$. Hence, we take into
account the following scattering potential at the junction

\begin{equation}
U(x)=(V_{1}-iV_{2})\delta(x).\label{eq:EffectH}
\end{equation}
This gives rise to the effective Hamiltonian $H_{\mathrm{eff}}=H_{S}+U(x)$. 

Explicitly, the Bogoliubov-de Gennes (BdG) Hamiltonian to describe
the NHJJ can be written as

\begin{equation}
\mathcal{H}_{\mathrm{BdG}}(x)=\left(\begin{array}{cc}
-\frac{\hbar^{2}\partial_{x}^{2}}{2m}-\mu+U(x) & \hat{\Delta}(x)\\
\hat{\Delta}^{\dagger}(x) & \frac{\hbar^{2}\partial_{x}^{2}}{2m}+\mu-U^{*}(x)
\end{array}\right),\label{eq:BdG}
\end{equation}
where $m$ is the effective mass of the electrons, $\mu$ the chemical
potential, and $\hat{\Delta}(x)$ the pairing potential. We argue
that the particle-hole symmetry (PHS) of the type $U\mathcal{H}_{\mathrm{BdG}}^{*}U^{-1}=-\mathcal{H}_{\mathrm{BdG}}$
is physically relevant in our model (see Appendix A), where $U$ is
a unitary matrix. Note that this particle-hole symmetry is labeled
as $\mathrm{PHS}^{\dagger}$ in the literature\ \cite{Kawabata19prl,Kawabata19prx}.\textcolor{blue}{{}
}Considering the energy spectrum of the system, electron and hole
excitations are described by the BdG equation

\begin{equation}
\mathcal{H}_{\mathrm{BdG}}(x)\psi(x)=E\psi(x),
\end{equation}
where the eigenstate $\psi(x)=(u(x),v(x))^{T}$ is a mixture of electron
and hole components, and $E$ is the excitation energy measured relative
to the Fermi energy. The eigenenergy of $\mathcal{H}_{\mathrm{BdG}}$
always comes in pairs $E\longleftrightarrow-E^{*}$ under $\mathrm{PHS}^{\dagger}$. 

\section{Andreev quasi-bound states and Josephson gaps}

We consider an $s$-wave NHJJ, in which the two superconductors are
both of $s$-wave type. Without loss of generality, we assume that
left and right superconductors have a constant pairing potential of
the same magnitude but different phases, i.e., $\hat{\Delta}(x<0)=\Delta$
and $\hat{\Delta}(x>0)=\Delta e^{i\phi}$ where $\phi$ is the phase
difference across the junction. We are interested in the low-energy
bound states existing within the superconducting gap. We obtain such Andreev quasi-bound states in the NHJJ by solving
the BdG equation with proper boundary conditions. We may write the
wave function of bound states as
\begin{equation}
\psi_{B}(x)=\sum_{\eta,\alpha}A_{\eta\alpha}e^{i\alpha\sigma_{\eta}k_{\eta}x}\left(\begin{array}{c}
e^{i\theta_{\eta\alpha}/2}\\
e^{-i\theta_{\eta\alpha}/2}
\end{array}\right),\label{eq:boundstates}
\end{equation}
where $\eta=e/h$, $\sigma_{e}\equiv1,$ $\sigma_{h}\equiv-1$, and
$\alpha=\pm$. The angles are defined as $\theta_{\eta+}\equiv\sigma_{\eta}\arccos(E_{B}/\Delta)+\phi$
and $\theta_{\eta-}\equiv\sigma_{\eta}\arccos(E_{B}/\Delta)$ with
the energy of bound states $E_{B}$. $A_{\eta\alpha}$ is the coefficient
of different evanescent modes. The bound states decay exponentially
for $|x|\rightarrow\infty$ with a finite decay length $\lambda$.
Therefore, the necessary condition for the existence of a bound state
is given by $\mathrm{Re}(\lambda)>0$ with $\lambda=\frac{\hbar v_{F}}{\Delta}\frac{1}{\sqrt{1-E_{B}^{2}/\Delta^{2}}}$,
where $v_{F}$ is the Fermi velocity. 

\subsection{Andreev quasi-bound states}

By considering the boundary conditions of $\psi_{B}(x)$ at $x=0$,
the secular equation to determine the energy of Andreev quasi-bound
state is given by 

\begin{equation}
(Z^{2}+1)+2iZ_{2}\sqrt{\frac{\Delta^{2}}{E_{B}^{2}}-1}=\frac{\Delta^{2}}{E_{B}^{2}}\left(Z^{2}+\cos^{2}\frac{\phi}{2}\right).\label{eq:E Basic}
\end{equation}
The parameter $Z_{j=1,2}$ for the strength of the non-Hermitian barrier
potential is defined as $Z_{j}\equiv\frac{mV_{j}}{\hbar^{2}k_{F}}$,
where $k_{F}$ is the Fermi wavevector and $Z^{2}\equiv\sum_{j}Z_{j}^{2}$. 

In the clean case, $Z_{1}=Z_{2}=0$, the above equation reduces to
the Kulik-Omel'yanchuk (KO) limit $E_{B}^{\pm}(\phi)=\pm\Delta\cos\frac{\phi}{2}$\ \cite{Kulik75jetps}.
For $Z_{1}\neq0$ but $Z_{2}=0$, we obtain the known result $E_{B}^{\pm}(\phi)=\pm\Delta\sqrt{\frac{Z_{1}^{2}+\cos^{2}\frac{\phi}{2}}{Z_{1}^{2}+1}}$\ \cite{Furusaki99sm}.
In this case, the Andreev spectrum has an energy gap $E_{g}=\frac{2\Delta Z_{1}}{\sqrt{Z_{1}^{2}+1}}$. 

Even if the JJ couples weakly to the environment $V_{2}\ll E_{g}$,
the imaginary term $2iZ_{2}\sqrt{\Delta^{2}/E_{B}^{2}-1}$ in Eq.~\eqref{eq:E Basic}
implies a complex Andreev spectrum, different from the Hermitian case\ \cite{Furusaki99sm,Beenakker91prl2,Kwon04epj,FuL09prb,Dolcini15prb,Beenakker13prl}.
The dimensionless parameter $Z_{2}$ can be estimated as $Z_{2}\sim\frac{V_{2}L}{\hbar v_{F}}$,
where $V_{2}$ describes the loss rate and $L$ is the junction length.
Considering reasonable numbers, $Z_{2}$ is estimated to be of the
order of $Z_{2}\sim0.1$\ \footnote{If we assume $V_{2}\sim1$ meV (a typical energy scale of our setup),
junction length $L\sim100$ nm, and Fermi velocity $v_{F}\sim1\times10^{6}$
m/s, the parameter $Z_{2}$ can be estimated as $Z_{2}\sim\frac{V_{2}L}{\hbar v_{F}}\approx0.1$.
The value of $Z_{2}$ can in principle be tuned by junction length
$L$ and loss rate $V_{2}$ by changing the coupling between JJ and
environment. }. We focus on this experimentally relevant regime of $Z_{2}\ll1$
below. 

At special values $\phi=2n\pi$ with integer $n$, a trivial solution
$E_{B}^{\pm}(\phi=2n\pi)=\pm\Delta$ exists, merging with the continuum.
At $\phi=(2n+1)\pi$, we find that the Andreev spectrum becomes

\begin{equation}
E_{B}^{+}[(2n+1)\pi]=\frac{Z^{2}\Delta}{\sqrt{Z^{4}+Z_{c}^{2}}},\label{eq:gapvalues}
\end{equation}
where $Z_{c}\equiv Z_{1}+iZ_{2}$ and $E_{B}^{-}=-(E_{B}^{+})^{*}$.
The real energy gap $E_{g}$ obtained from Eq.~\eqref{eq:gapvalues}
is illustrated in Fig. \ref{fig1:setup}(c). Evidently, the energy
gap $E_{g}$ is just slightly modified as we turn on a small $Z_{2}$
in the weak coupling regime. 

For a general phase difference $\phi\in[\phi_{b},\phi_{t}]$, the
secular equation can be written as
\begin{widetext}
\begin{equation}
\frac{E^{2}}{\Delta^{2}}=\frac{Z^{2}(Z^{2}-\sin^{2}\frac{\phi}{2})+\cos^{2}\frac{\phi}{2}+2Z_{1}^{2}\pm2Z_{2}\sqrt{\cos^{2}\frac{\phi}{2}(Z^{2}-\sin^{2}\frac{\phi}{2})-Z_{1}^{2}}}{(Z^{2}+1)^{2}-4Z_{2}^{2}}.\label{eq:ABS s1}
\end{equation}
\end{widetext}

We plot the corresponding complex Andreev spectrum obtained from the
above equation in Fig. \ref{fig1:setup}(b).\textcolor{blue}{{} }The
real part of the complex Andreev spectrum indicates the physical energy
while its imaginary part characterizes a finite lifetime of quasiparticles
traveling across the junction. The phase boundaries are defined as
$\phi_{b}\equiv2n\pi+\phi_{0}(Z_{2})$ and $\phi_{t}\equiv(2n+1)\pi-\phi_{0}(Z_{2})$
with $\phi_{0}(Z_{2})\equiv2\arcsin(\sqrt{2}Z_{2}/\sqrt{1+Z^{2}})$
obtained from the condition $\mathrm{Re}(\lambda)>0$. For $Z_{2}\ll1$,
this phase boundary can be approximated by $\phi_{0}(Z_{2})\simeq\frac{2\sqrt{2}Z_{2}}{\sqrt{1+Z_{1}^{2}}}$.

\subsection{Josephson gaps}

\begin{figure}
\includegraphics[width=1\linewidth]{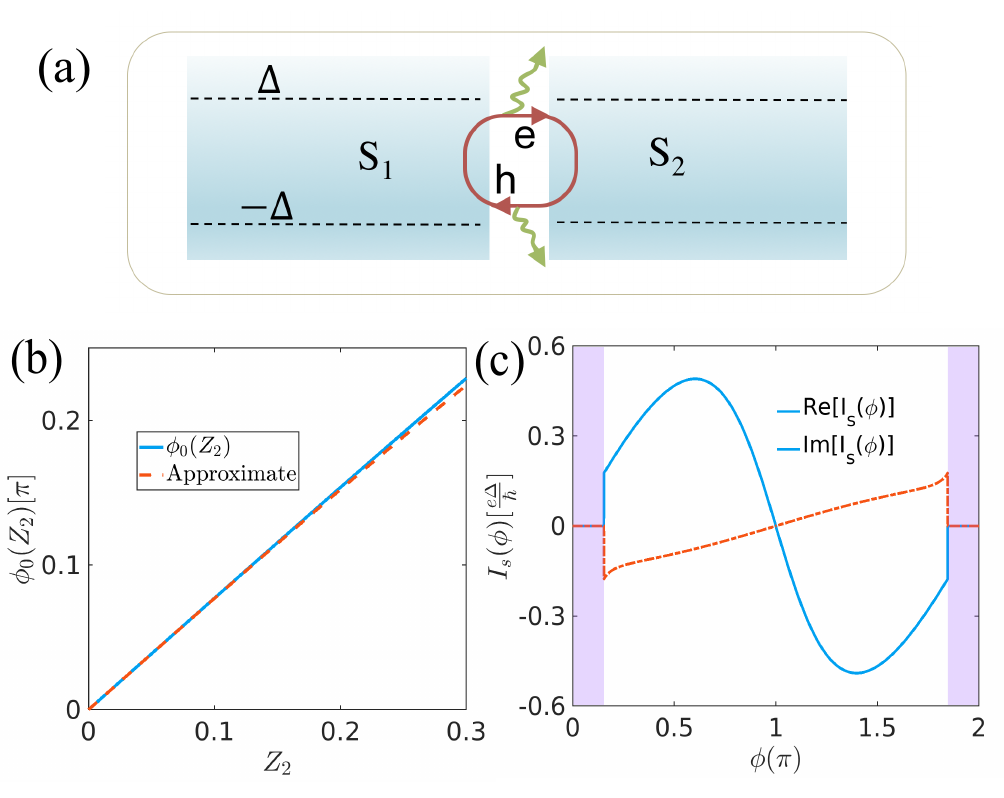}

\caption{(a) Sketch of inelastic Cooper pair tunneling. Transfer of Cooper
pairs experiences possible decoherence and quasiparticles loss (wavy
lines). (b) Josephson-gap phase boundary $\phi_{0}(Z_{2})$ as a function
of $Z_{2}$. (c) Complex supercurrents as a function of $\phi$ with
$Z_{1}=0.6$ and $Z_{2}=0.2$. \label{fig2:supercurrent}}
\end{figure}

In general, the Andreev spectrum is a continuous function with respect
to the phase difference $\phi$. However, we find that within a finite
phase window $\Phi_{W}\equiv[2n\pi-\phi_{0}(Z_{2}),2n\pi+\phi_{0}(Z_{2})]$
with $n\in\mathbb{Z}$, low-energy Andreev quasi-bound states are
not allowed at all within the superconducting gap $\Delta$. We call
such phase windows Josephson gaps. Its gap value is $2\phi_{0}(Z_{2})=4\arcsin(\sqrt{2}Z_{2}/\sqrt{1+Z^{2}})$.
The appearance of Josephson gaps can be understood as a consequence
of the competition between phase-coherent transport of Cooper pairs
across the junction and decoherence induced by the NH barrier. This
competition can be understood by comparing the phase difference $\phi$
with the loss parameter $Z_{2}$. In the Hermitian case, due to phase
coherence across the junction, quasiparticles move back and forth
in the junction carrying the phase factor of superconductors imprinted
by Andreev reflection\ \cite{Tinkham,Nazarovbook,Golubov04rmp,Furusaki99sm}.\textcolor{magenta}{{}
} Andreev reflections lead to phase-coherent transport of Cooper pairs
across the junction, i.e. dc supercurrent. However, the NH mechanism
in the NHJJ gives rise to loss of coherence and inelastic Andreev
reflections {[}see Fig. \ref{fig2:supercurrent}(a){]}\ \cite{Nazarovbook,Daley14AIP,Minganti19pras}.
The magnitude of $Z_{2}$ determines the strength of decoherence.
Hence, it is clear that the larger $Z_{2}$, the stronger is the coupling
between system and environment. For phase differences $\phi\sim0$
(mod $2\pi$) associated with a small supercurrent meaning a small
number of transferred Cooper pairs, we observe Josephson gaps. This
implies that phase-coherent transport of Cooper pairs is totally absent
(thus no bound states). For larger supercurrents, i.e. larger phase
differences, the pair breaking mechanism related to the barrier strength
$Z_{2}$ is not sufficient to block the supercurrent. Nevertheless,
decoherence modifies the supercurrent in this case. The Josephson-gap
phase boundary $\phi_{0}(Z_{2})$ depends on the value of $Z_{2}$.
In Fig. \ref{fig2:supercurrent}(b), we plot the Josephson-gap phase
boundary as a function of $Z_{2}$. It grows almost linearly with
increasing $Z_{2}$. 

\section{Inelastic Cooper pair tunneling and complex supercurrent}

In (short) Hermitian JJs, the supercurrent originates from the coherent
Cooper pair tunneling related to the Andreev spectrum\ \cite{Beenakker91prl,Beenakker92proceed,Furusaki99sm}.
In NHJJs, the Andreev spectrum becomes complex. The imaginary part
of the Andreev spectrum indicates a finite lifetime of quasiparticles.
Thus, the NH mechanism introduces loss of quasiparticles and decoherence
during the Andreev reflection processes\ \cite{Breuer02book,Nazarovbook,Daley14AIP,Minganti19pras}
{[}see Fig. \ref{fig2:supercurrent}(a){]}, leading to inelastic Cooper
pair tunneling \ \cite{Devoret90prl,Nazarov94proceed,Holfheinz11prl,Westig17prl,Grimm19prx}. 

The supercurrent $I_{s}(\phi)$ can be obtained by taking the derivative
of the free energy with respect to the phase difference $\phi$\ \cite{Beenakker92proceed}.
It leads to $I_{s}(\phi)=-\frac{2e}{\hbar}\sum_{n>0}\frac{dE_{n}(\phi)}{d\phi}$
with the Andreev spectrum $E_{n}(\phi)$ in Hermitian JJs\ \cite{Beenakker91prl,Beenakker92proceed}.
In NHJJ, the real part of the Andreev spectrum is the physical energy
while the imaginary part corresponds to a finite lifetime of each
eigenstate. The free energy in the ground state of NHJJs can be interpreted
as the summation of real energy levels, broadened due to imaginary
parts up to the Fermi energy\ \cite{Yamamoto19prl}. Following the
same reasoning as in Hermitian JJs, we derive the supercurrent as

\begin{equation}
I_{s}(\phi)=-\frac{2e}{\hbar}\frac{d\mathrm{Re}[E_{B}^{+}(\phi)]}{d\phi}+i\frac{2e}{\hbar}\frac{d\mathrm{Im}[E_{B}^{+}(\phi)]}{d\phi},
\end{equation}
where $E_{B}^{+}(\phi)$ is the complex Andreev spectrum. The real
part $\mathrm{Re}[I_{s}(\phi)]$ represents the physical supercurrent
flowing across JJs. The imaginary part $\mathrm{Im}[I_{s}(\phi)]$
stems from phase-dependent lifetimes of quasiparticles during inelastic
Cooper pair tunneling. Thus, it is associated with the loss of quasiparticles
from the JJ to the environment \footnote{The lifetime of quasiparticles becomes phase dependent. Thus, the
magnitude of imaginary supercurrent indicates the response of quasiparticles
loss as changing $\phi$.}. Indeed, it is possible for either electrons or holes to escape the
JJs after experiencing inelastic scattering at the NH barrier. If
we consider scattering of electrons or holes, the scattering probability
becomes $P(E,Z_{1},Z_{2})<1$, which means that some quasiparticles
escape into the environment from the JJ (see Appendix B). 

Explicitly for $s$-wave NHJJs, within the phase region $\phi\in[\phi_{b},\phi_{t}]$,
the supercurrent as a function of $\phi$ is plotted in Fig. \ref{fig2:supercurrent}(c).
Within the Josephson gaps, there is no supercurrent flowing across
the junction. The supercurrent exhibits a jump at the phase boundary
$\phi_{0}(Z_{2})$.\textcolor{blue}{{} }We note that imaginary currents
in normal states have been employed to describe delocalized behavior
or dissipation of eigenstates\ \cite{Hatano96prl,ZhangSB22prb,LiQ22prb},
while to the best of our knowledge, a complex supercurrent has not
been addressed before. 

\section{Experimental considerations and numerical simulations }

Our analysis of NHJJs is experimentally relevant, considering the
advanced fabrication techniques of JJs. The $s$-wave JJs could, for
instance, be built by aluminum-based or InAs-based superconductor
nanowires\ \cite{Doh05science,Chang15nn,Hays18prl}. The local loss
at the junction may stem from the coupling of JJs to the environment
with a dissipative lead\ \cite{ZhangS22prl,LiuD22prl}. In the following,
we provide more evidence from numerical simulations.

To perform tight-binding calculations, we discretize the Hamiltonian
on a one-dimensional lattice by the long-wavelength approximation.
The lattice constant is set at $a=1$. We divide the system into three
parts: the left side and the right side are superconductors of length
$L_{s}$, and the middle region is the normal part of length $L_{N}$.
To mimic the loss at the junction, we set $L_{N}=1a$ and an on-site
complex potential $V_{1}-iV_{2}$. The two superconductors have a
paring potential $\Delta$ and a phase difference $\phi$. 

\begin{figure}
\includegraphics[width=1\linewidth]{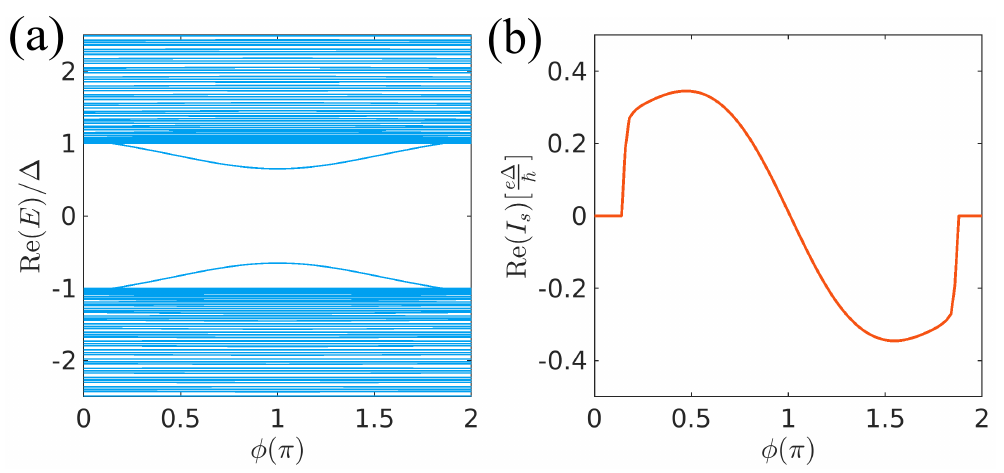}

\caption{Tight-binding simulations of the $s$-wave NHJJ with a finite length.
(a) Real part of energy spectra as a function of $\phi$. (b) Real
supercurrent as a function of $\phi$ obtained from corresponding
energy spectra in (a). We choose $V_{1}=0.8$, $V_{2}=0.3$,
$\mu=1$ in units of $\Delta$, and $L_{s}=100a$ here. \label{figTBSwave}}
\end{figure}

Figure \ref{figTBSwave}(a) shows the real part of the Andreev spectra
as a function of $\phi$ for the NHJJ. Note the presence of the Josephson
gap. Figure \ref{figTBSwave}(b) illustrates the real part of the
supercurrent obtained from the Andreev spectra corresponding to Fig.
\ref{figTBSwave}(a). The supercurrent vanishes within the Josephson
gap. 

To evaluate the phase boundary $\phi_{0}(Z_{2})$, we need the estimate
$Z_{j}$ via $Z_{j}=\frac{mV_{j}}{\hbar^{2}k_{F}}$. The phase boundary
is then obtained as $\phi_{0}(Z_{2})=2\arcsin\left(\frac{\sqrt{2}Z_{2}}{\sqrt{1+Z^{2}}}\right)$.
For the parameters used in Fig. \ref{figTBSwave}(b), we obtain $Z_{1}=0.566$,
and $Z_{2}=0.212$, which gives $\phi_{0}(Z_{2})=0.165\pi$, in accordance
with the plot.

\section{discussion and conclusions}

To summarize, we have studied the low-energy spectral features and
transport properties of NHJJs. Our minimal model of NHJJs reveals
anomalous NH physics emerging in JJs. We have introduced several conceptually
new phenomena, including Josephson gaps and complex supercurrents,
with no counterparts in the Hermitian case. 

Our work extends the realm of the Josephson effect to NH physics and
inspires potential applications: the NHJJ may function as a supercurrent
switch by tuning the phase difference $\phi$ exploiting the presence
of Josephson gaps. The impact of non-Hermiticity on other related
phenomena, such as ac Josephson effects, Shapiro steps\ \cite{Tinkham},
JJ circuits\ \cite{Devoret13science}, and Josephson diode effects\ \cite{Ando20nature},
are interesting extensions of our work. 

\section{Acknowledgments}

We thank Jan Budich, Haiping Hu, Jian Li, Zhuo-Yu Xian, and Chunxu
Zhang for valuable discussions. This work was supported by the Würzburg-Dresden
Cluster of Excellence ct.qmat, EXC2147, project-id 390858490, the
DFG (SFB 1170), and the Bavarian Ministry of Economic Affairs, Regional
Development and Energy within the High-Tech Agenda Project \textquotedblleft Bausteine
für das Quanten Computing auf Basis topologischer Materialen\textquotedblright .

\textit{Note added.---}Recently, we came to notice a related work
on NHJJs\ \cite{Cayao23arxiv}.

\appendix

\section{Relevant particle-hole symmetry of the model}

In this section, we analyze the particle-hole symmetry that is physically
relevant from a general form. From a field operator perspective, the
particle-hole symmetry (PHS) mixes creation operator $\Psi^{\dagger}$
and annihilation operator $\Psi$ in a way 

\begin{equation}
C\Psi_{\alpha}C^{-1}=U_{\alpha\beta}\Psi_{\beta}^{\dagger},\ C\Psi_{\alpha}^{\dagger}C^{-1}=\Psi_{\alpha}U_{\alpha\beta}^{T},\label{eq:PHoperator}
\end{equation}
where $U$ represents a unitary matrix and $C$ denotes the PHS operator.
A Hamiltonian $H=\Psi_{\alpha}^{\dagger}\mathcal{H}_{\alpha\beta}\Psi_{\beta}$
is particle-hole symmetric if $CHC^{-1}=H$. This implies that the
first-quantized Hamiltonian transforms under PHS as
\begin{equation}
C\mathcal{H}^{T}C^{-1}=-\mathcal{H},
\end{equation}
where $T$ is the transpose operation. In the Hermitian case, $\mathcal{H}^{T}=\mathcal{H}^{*}$.
While in the non-Hermitian case, $\mathcal{H}^{T}\neq\mathcal{H}^{*}$
in general. Thus, the PHS comes in two kinds as\ \cite{Kawabata19prl}:

\begin{align}
U\mathcal{H}^{*}U^{-1} & =-\mathcal{H},\ \text{\ensuremath{\mathrm{type}}\ I};\nonumber \\
U\mathcal{H}^{T}U^{-1} & =-\mathcal{H},\ \mathrm{\text{\ensuremath{\mathrm{type}}\ I}I.}\label{eq:symmetry}
\end{align}
In the presence of PHS, the excitation energy of $\mathcal{H}$ always
comes in pairs. For the two types of PHSs above, the corresponding
energy pairs are $E\longleftrightarrow-E^{*}$ for type I PHS and
$E\longleftrightarrow-E$ for type II PHS, respectively. 

To determine which one of the two types of PHS on $\mathcal{H}$ is
physically more relevant, we further consider the constraint of PHS
on Green's functions. On the one hand, the effective Hamiltonian $\mathcal{H}$
is directly related with the retarded Green's function. Note that
the poles of the retarded Green's function yield the eigenvalues of
the Hamiltonian, located in the lower half of the complex plane. On
the other hand, the physical indication of the retarded Green's function
as a propagator is consistent with causality. The retarded Green's
function is defined in terms of field operators as $G_{\alpha\beta}^{R}(t)=-i\theta(t)\langle[\Psi_{\alpha}(t),\Psi_{\beta}^{\dagger}(0)]\rangle$,
where $t$ is the time domain and $\theta(t)$ is a Heaviside step
function. Considering PHS on field operators in Green's function,
it leads to
\begin{alignat}{1}
G_{\alpha\beta}^{R}(t) & =-i\theta(t)\langle[\Psi_{\alpha}(t),\Psi_{\beta}^{\dagger}(0)]\rangle\nonumber \\
 & =-i\theta(t)\langle C^{-1}C[\Psi_{\alpha}(t),\Psi_{\beta}^{\dagger}(0)]\rangle\nonumber \\
 & =-i\theta(t)\langle C[\Psi_{\alpha}(t),\Psi_{\beta}^{\dagger}(0)]C^{-1}\rangle\nonumber \\
 & =-U_{\alpha\gamma}^{*}\langle i\theta(t)[\Psi_{\gamma}^{\dagger}(0),\Psi_{\delta}(-t)]\rangle U_{\delta\beta}^{T}\nonumber \\
 & =-U_{\alpha\gamma}^{*}G_{\gamma\delta}^{A}(-t)U_{\delta\beta}^{T}.
\end{alignat}
By Fourier transformation and relating the retarded and advanced Green's
function by $[G_{\alpha\beta}^{R}(E)]^{*}=G_{\alpha\beta}^{A}(E)$,
the retarded Green's function fulfills

\begin{equation}
U^{\dagger}[G^{R}(E)]^{*}U=-G^{R}(-E).
\end{equation}
This PHS constraint on retarded Green's function is consistent with
type I PHS acting on $\mathcal{H}$: The eigenvalues of the system
(poles of the Green's function) reside on the same side of the complex
plane $\mathrm{Im}[E]\leq0$. In contrast, from the constraint of
type II PHS acting on $\mathcal{H}$, eigenvalues distribute symmetrically
in the upper and lower half of the complex plane. This symmetric eigenenergy
distribution is not consistent with the eigenvalues obtained from
the retarded Green's function. Thus, it is not compatible with the
requirement of causality. Therefore, we argue that type I PHS acting
on $\mathcal{H}$ is physically more relevant to describe non-Hermitian
systems. This type I PHS is employed in the main text and labeled
as $\mathrm{PHS^{\dagger}}$. 

\section{Quasiparticle loss due to non-Hermitian barrier scattering}

In this section, we obtain the reflection amplitudes (Andreev reflection
and the normal reflection) and tunneling amplitudes (crossed Andreev
tunneling and the normal tunneling) of quasiparticles after scattering
at the non-Hermitian barrier. We show explicitly the loss of quasiparticles
from the junction to the environment due to non-Hermitian scatterings.
Consider an electron-like quasiparticle that is propagating from the
left side of the junction. The wave function can be written as 
\begin{widetext}
\begin{equation}
\psi(x)=\left(\begin{array}{c}
u(x)\\
v(x)
\end{array}\right)=\begin{cases}
e^{ik_{e}x}\left(\begin{array}{c}
u_{0}\\
v_{0}
\end{array}\right)+A_{h-}e^{ik_{h}x}\left(\begin{array}{c}
v_{0}\\
u_{0}
\end{array}\right)+A_{e-}e^{-ik_{e}x}\left(\begin{array}{c}
u_{0}\\
v_{0}
\end{array}\right), & x<0;\\
A_{e+}e^{ik_{e}x}\left(\begin{array}{c}
u_{0}e^{i\phi/2}\\
v_{0}e^{-i\phi/2}
\end{array}\right)+A_{h+}e^{-ik_{h}x}\left(\begin{array}{c}
v_{0}e^{i\phi/2}\\
u_{0}e^{-i\phi/2}
\end{array}\right), & x>0.
\end{cases}
\end{equation}
Note that $u_{0}^{2}=\frac{1}{2}\left(1+i\frac{\sqrt{\Delta^{2}-E^{2}}}{E}\right),\ v_{0}^{2}=\frac{1}{2}\left(1-i\frac{\sqrt{\Delta^{2}-E^{2}}}{E}\right)$.
The wave vectors are defined as $\hbar k_{e}=\sqrt{2m(\mu+i\sqrt{\Delta^{2}-E^{2}})},\ k_{h}=k_{e}^{*}.$
The coefficients are determined by the boundary conditions:
\begin{align}
\psi(0+) & =\psi(0-),\\
\psi'(0+)-\psi'(0-) & =\frac{2m}{\hbar^{2}}(V_{1}\tau_{0}-iV_{2}\tau_{z})\psi(0+).
\end{align}
By solving these linear equations, the four coefficients (Andreev
reflection $A_{h-},$ normal reflection $A_{e-}$, normal tunneling
$A_{e+}$, crossed Andreev tunneling $A_{h+}$) are obtained as 
\begin{align}
A_{h-} & =-\frac{\Delta[E\sin^{2}\frac{\phi}{2}+\Omega(i\sin\frac{\phi}{2}\cos\frac{\phi}{2}-Z_{2})]}{(Z^{2}+1)E^{2}+2Z_{2}\Omega E-\Delta^{2}(Z^{2}+\cos^{2}\frac{\phi}{2})},\\
A_{e-} & =-\frac{Z^{2}\Omega^{2}-Z_{2}\Omega E-iZ_{1}\Omega^{2}}{(Z^{2}+1)E^{2}+2Z_{2}\Omega E-\Delta^{2}(Z^{2}+\cos^{2}\frac{\phi}{2})},\\
A_{e+} & =-\frac{\Omega E[Z_{2}\cos\frac{\phi}{2}+i(1+iZ_{1})\sin\frac{\phi}{2}]-\Omega^{2}[(1+iZ_{1})\cos\frac{\phi}{2}+iZ_{2}\sin\frac{\phi}{2}]}{(Z^{2}+1)E^{2}+2Z_{2}\Omega E-\Delta^{2}(Z^{2}+\cos^{2}\frac{\phi}{2})},\\
A_{h+} & =\frac{\Delta\Omega[Z_{2}\cos\frac{\phi}{2}-Z_{1}\sin\frac{\phi}{2}]}{(Z^{2}+1)E^{2}+2Z_{2}\Omega E-\Delta^{2}(Z^{2}+\cos^{2}\frac{\phi}{2})}.
\end{align}
\end{widetext}

\begin{figure}
\includegraphics[width=0.8\linewidth]{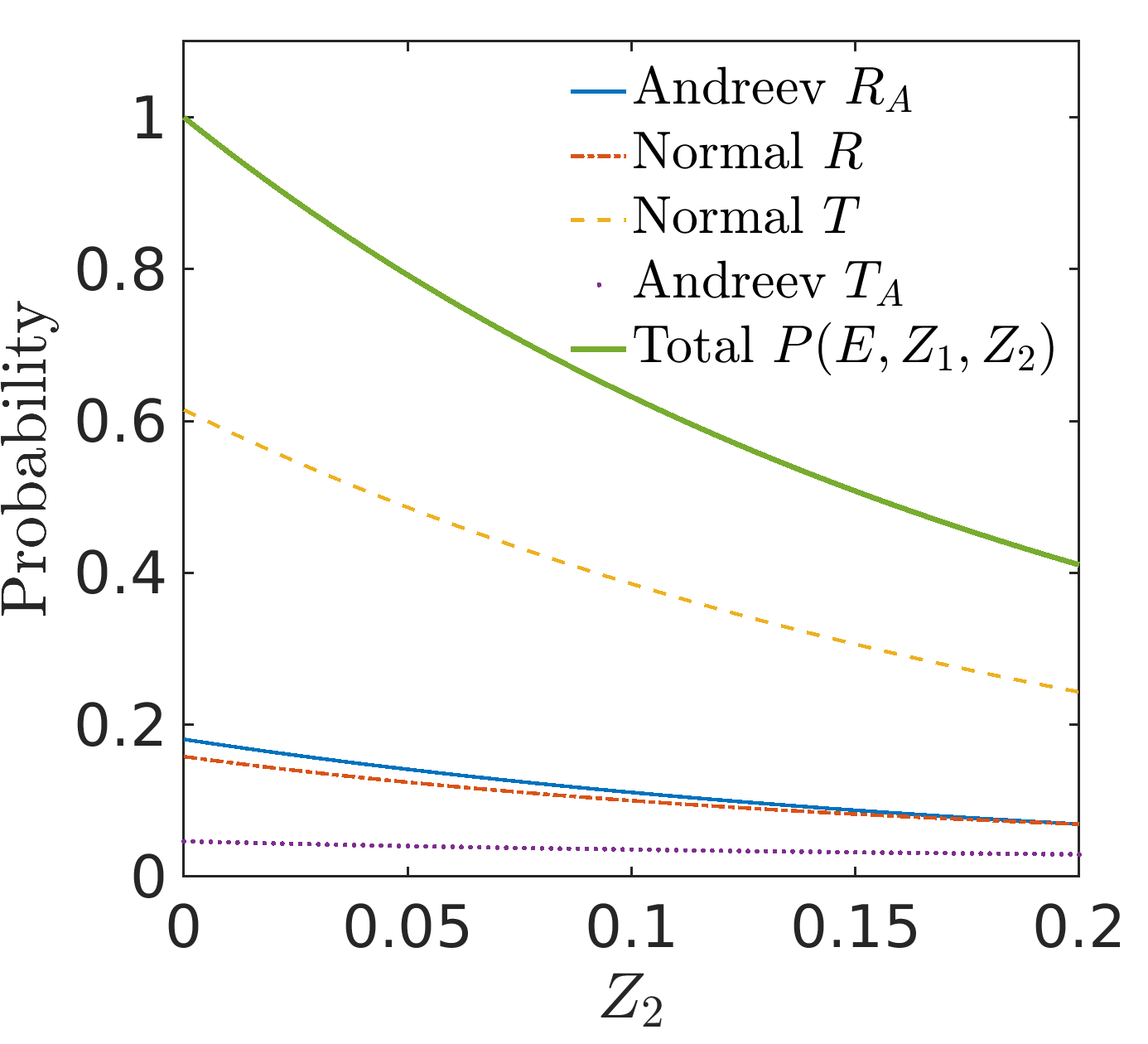}

\caption{Scattering probabilities as a function of $Z_{2}$ for fixed $Z_{1}=0.6$.
We chose $\phi=\frac{\pi}{2}$ in this plot. We choose $E=1.5$ in
unit of $\Delta$. \label{figScatteringP}}
\end{figure}

The parameter $\Omega$ is defined as $\Omega\equiv\sqrt{E^{2}-\Delta^{2}}$.
The total scattering probability is defined as 

\begin{equation}
P(E,Z_{1},Z_{2})=\sum_{\eta=e/h,s=\pm}|A_{\eta s}|^{2}.
\end{equation}
In the Hermitian case, the probability is conserved at $P(E,Z_{1},Z_{2}=0)=1$.
For nonzero $Z_{2}$, we find that the probability $P(E,Z_{1},Z_{2})<1$,
which indicates the loss of quasiparticles from the junction to the
environment {[}as shown in Figs. \ref{figScatteringP}(a) and \ref{figScatteringP}(b){]}.
The probability $P(E,Z_{1},Z_{2})$ decays continuously as increasing
the strength of $Z_{2}$. Thus, the value of $Z_{2}$ measures the
magnitude of loss.

\end{document}